\documentclass[12pt]{article} 
\usepackage{amsmath,amssymb}
\newcommand{\beq}{\begin{eqnarray}}
\newcommand{\eeq}{\end{eqnarray}}

\newcommand{\qed}{{\it Q.E.D.\/} \bigskip\par}

\newcommand{\rd}{\partial}

\renewcommand{\Re}{\mathrm{Re}\,}
\newcommand{\bbC}{\mathbb{C}}
\newcommand{\bbP}{\mathbb{P}}

\newcommand{\bbZ}{\mathbb{Z}}

\newcommand{\calL}{\mathcal{L}}

\newcommand{\Res}{\operatorname{Res}}

\newcommand{\SU}{\mathrm{SU}}
\newcommand{\sh}{\operatorname{sinh}\nolimits}
\newcommand{\ch}{\operatorname{cosh}\nolimits}
\newcommand{\cth}{\operatorname{coth}\nolimits}
\newcommand{\commentout}[1]{}

\begin{document}

\title{An integrable system \\
on the moduli space of rational functions\\
and its variants}
\author{Kanehisa Takasaki\\
\normalsize Department of Fundamental Sciences, \\
\normalsize Faculty of Integrate Human Studies, Kyoto University\\
\normalsize Yoshida, Sakyo-ku, Kyoto 606-8501, Japan\\
\normalsize E-mail: takasaki@math.h.kyoto-u.ac.jp\\
\normalsize and\\
Takashi Takebe\\
\normalsize Department of Mathematics, \\
\normalsize Faculty of Science, Ochanomizu University\\
\normalsize Otsuka, Bunkyo-ku, Tokyo 112-8610, Japan\\
\normalsize E-mail: takebe@math.ocha.ac.jp}
\date{}
\maketitle

\begin{abstract}
 We study several integrable Hamiltonian systems on the moduli spaces of
 meromorphic functions on Riemann surfaces (the Riemann sphere, a cylinder
 and a torus). The action-angle variables and the separated variables
 (in Sklyanin's sense) are related via a canonical transformation, the
 generating function of which is the Abel-Jacobi type integral of the
 Seiberg-Witten differential over the spectral curve.
\end{abstract}

\vfill


\begin{flushleft}
nlin.SI/0202042
\end{flushleft}

\newpage

\section{Introduction}

Moser's work \cite{bib:Mo-1} on the open finite Toda lattice 
will be presumably the earliest attempt in the literature to 
relate a moduli space of rational functions to an integrable 
system.  The idea is to construct a rational function 
\begin{equation*}
  f(\lambda) = \sum_{j=1}^N \frac{\rho_j}{\lambda - \alpha_j} 
\end{equation*}
from the $L$-matrix of the Toda lattice, which can be 
reproduced from $f(\lambda)$ by a continued fraction.  
Moser discovered that the dynamics of the Toda lattice is 
linearized in the new variables $\rho_j$ and $\alpha_j$ 
(i.e., moduli of the rational function), so that an explicit 
formula of solutions of the Toda lattice can be obtained 
by inverting the map $L \to f(\lambda)$.  The mechanism of 
linearization was elucidated (and generalized) by Kostant 
\cite{bib:Ko} in the language of representations of Lie groups.  

It was soon noticed that Moser's method resembles 
mathematical techniques of the system control theory 
in engineering.  An analogue of Moser's rational function 
emerges therein as a fundamental quantity (the rational 
transfer function) that characterizes the input-output relation 
of a linear control system.  Following Moser's construction, 
Krishnaprasad \cite{bib:Kr} introduced dynamical flows on 
the moduli space of such rational functions, and solved 
some geometrical problems raised by Brockett \cite{bib:Br}.  
Krishnaprasad's work was further refined by Nakamura 
\cite{bib:Na-1}, who thus obtained a few variants of 
the open finite Toda lattice.  

Another notable example where a moduli space of 
rational functions plays a role, is the monopole 
moduli space of the $\SU(2)$ Yang-Mills theory. 
According to Donaldson's result \cite{bib:Do}, 
the monopole moduli space is isomorphic to a space of 
rational functions (though with complex coefficients, 
as opposed to Moser's case).  It is not difficult to 
infer the existence of an integrable system on this 
moduli space from Donaldson's work (and the subsequent 
progress in the book of Atiyah and Hitchin \cite{bib:At-Hi}).  
Firstly, although Donaldson mentioned nothing, 
Donaldson's construction of the rational function 
is actually very similar to the rational transfer 
function in the system control theory.  Secondly, 
Atiyah and Hitchin \cite{bib:At-Hi} introduced 
a symplectic structure on Donaldson's space of 
rational functions, which can be the first step 
towards the construction of an integrable system.  
{}From the first point of view, Nakamura \cite{bib:Na-2} 
extended his previous work \cite{bib:Na-1} to 
this space of complex rational functions. 
The second point of view was sought by Faybusovich 
and Gekhtman \cite{bib:Fa-Ge-1}, who recently reported 
the existence of a multi-Hamiltonian structure therein 
\cite{bib:Fa-Ge-2}.  

In this paper, we take up this relatively well understood 
integrable system from two new aspects: 
\begin{enumerate}
\item 
{\em Action-angle variables of Seiberg-Witten type\/}. 
Needless to say, the existence of action-angle variables 
is the most fundamental aspect of integrable systems.  
The so called Seiberg-Witten integrable systems 
\cite{bib:Ma-Wa,bib:Do-Wi,bib:Do-SW} are characterized 
by a special set of action-angle variables that are 
associated with a special differential $dS$ (the 
Seiberg-Witten differential) on the spectral curve.  
We shall show that the integrable system on the moduli 
space of rational functions has a similar property.  
\item 
{\em Separation of variables\/}.  Another remarkable 
aspect of many finite-dimensional integrable systems 
is separation of variables (SoV).  In the modern version 
of SoV due to Sklyanin \cite{bib:Sk}, an integrable system 
is mapped to a dynamical system of a finite number of 
points on the spectral curve, and the coordinates of 
these moving points are nothing but separated variables.  
(A prototype of this interpretation of SoV can be found 
in Moser's work on classical integrable systems 
\cite{bib:Mo-2} and its generalization by the Montreal 
group \cite{bib:AHH-SOV}.)  The integrable system 
on the moduli space of rational functions turns out to 
have a similar set of separated variables.
\end{enumerate}
The role of the spectral curve is now played by 
a {\it rational curve} of the form 
\begin{equation*}
    C = \{(\lambda,z) \mid z = A(\lambda)\}, 
\end{equation*}
where $A(\lambda)$ is the denominator of the expression 
\begin{equation*}
  f(\lambda) = \frac{B(\lambda)}{A(\lambda)} 
\end{equation*}
of $f(\lambda)$ as the quotient of two polynomials. 
The separated variables are given by the zeros $\lambda_j$ 
of $B(\lambda)$ and the values $z_j = A(\lambda_j)$ of 
$A(\lambda)$ at these zeros.  In the context of 
Seiberg-Witten integrable systems, this kind of rational 
spectral curves emerge in the weak coupling limit of the 
supersymmetric gauge theory \cite{bib:Br-etal-5D,bib:Br-Ma}.  

We construct two variants of this integrable system 
on the basis of these observations.  The new integrable 
systems live on a moduli space of trigonometric or 
elliptic functions, and inherit most of the properties 
of the rational case.   We show that these integrable 
systems can be formulated in a unified way, which will be 
useful for higher genus generalization. 

Further generalization can be sought in the perspective 
of complex surfaces with symplectic or Poisson structure 
\cite{bib:Hu,bib:Va,bib:Bo}.  All the three 
(rational, trigonometric and elliptic) cases are 
associated with a complex surface $X$ with rational 
fibration over a (rational or elliptic) curve $\Sigma$. 
The curve $C$ is embedded in $X$.  A natural idea of 
generalization is to replace the rational fibration by 
elliptic fibration.  A few examples of that kind of 
integrable systems have been constructed \cite{bib:Ta-K3}.  
We reexamine these examples in the present context.  

A few more comments on the rational case are in order. 
Firstly, Morosi and Tondo \cite{bib:Mo-To} considered 
a very similar example of separation of variables.  
Their example is picked out from Calogero's many-body systems 
describing motion of zeros (or poles) of solutions of a partial 
differential equation \cite{bib:Ca}.  This example, too, 
can be reformulated as an integrable system on the moduli 
space of rational functions except that the degree of 
the denominator $B(\lambda)$ is different from our case.  
Secondly, Krichever and Vaninsky \cite{bib:Kr-Va} developed 
an algebro-geometric approach to the finite open Toda lattice 
using Baker-Akhiezer functions on a singular rational curve.  
The Seiberg-Witten structure and separated variables can be 
reproduced from their results as well.  

This paper is organized as follows.  
In Section \ref{sec:out}, we show an outline of 
our construction.  Section \ref{sec:rat} reviews 
the case of rational functions from the new points 
of view.  Integrable systems on a space of trigonometric 
and elliptic functions are constructed in Sections 
\ref{sec:tri} and \ref{sec:ell}.  Section \ref{sec:sur} 
seeks generalizations in the context of symplectic 
surfaces.  We show our conclusion in Section \ref{sec:conc}.

\section{Outline of construction}
\label{sec:out}

Our strategy of constructing the integrable systems is summarized in the
following way. We do not consider global nature of the phase space and
restricted ourselves to generic situation.
\begin{enumerate}
 \item The phase space is the moduli space of meromorphic functions of
       certain class on a fixed Riemann surface $\Sigma$, which is
       either a sphere, a cylinder or a torus.  We
       factorize each meromorphic function as $B(\lambda)/A(\lambda)$,
       where $A(\lambda)$ and $B(\lambda)$ are holomorphic functions (or
       holomorphic sections of a line bundle) on $\Sigma$.

 \item Let $\{\alpha_j\}$ be the set of poles of the meromorphic
       function $B(\lambda)/A(\lambda)$, i.e., zeros of $A(\lambda)$. We
       have a canonical coordinate system on the phase space $(\alpha_j,
       \psi_j)_{j=1,\dots,N}$. Hence the canonical $2$-form $\Omega$ of
       the phase space has the form
\begin{equation}
    \Omega = \sum_{j=1}^N d\alpha_j \wedge d\psi_j.
\end{equation}

 \item The Hamiltonians of the system $u_n(\alpha)$ ($n=1,\dots,N$) are
       given. Not containing the variables $\psi_j$, they mutually
       commute; $\{u_n, u_m\} = 0$. By coordinate transformation, we can
       rewrite $\Omega$ in the form
\begin{equation}
    \Omega = \sum_{n=1}^N du_n \wedge d\phi_n.
\end{equation}
       Namely $\phi_n$ is the angle variable corresponding to the action
       variable $u_n$. In this sense, the system is trivially solved.

 \item There exists another set of canonical variables $(\lambda_j,
       \mu_j)_{j=1,\dots,N}$ for which $\Omega$ can be written as 
\begin{equation}
    \Omega = \sum_{j=1}^N d\mu_j \wedge d\lambda_j,  
\end{equation}
       where $\{\lambda_j\}$ is the set of zeros
       of $B(\lambda)/A(\lambda)$, i.e., zeros of $B(\lambda)$. Each
       pair $(\lambda_j, z_j = e^{\mu_j})$ ($j=1,\dots,N$) satisfies a
       relation,
\begin{equation}
    z_j = A(\lambda_j; u_1,\dots, u_N).
\label{sov-eq}
\end{equation}

 \item The generating function $S(\lambda_1,\dots,\lambda_N;
       u_1,\dots,u_N)$ of the canonical transformation
\begin{equation}
    (\lambda_j, \mu_j)_{j=1,\dots,N} 
    \mapsto (u_n, \phi_n)_{n=1,\dots,N}
\end{equation}
       is defined by the integral
\begin{equation}
    S(\lambda_1,\dots,\lambda_N;u_1,\dots,u_N)
    =
    \sum_{i=1}^N \int^{\lambda_i} 
    \log A(\lambda;u_1,\dots,u_N)\, d\lambda.
\label{sov-gen-func}
\end{equation}
        Hence the variables $\mu_j$ and $\phi_n$ are expressed as
\begin{equation}
 \begin{split}
    \mu_j &= \frac{\rd S}{\rd \lambda_j},
\\
    \phi_n &= \frac{\rd S}{\rd u_n}
    =\sum_{i=1}^N 
    \int^{\lambda_i} 
      \frac{\rd A}{\rd u_n}(\lambda;u_1,\dots,u_N)\, 
      \frac{d\lambda}{A(\lambda;u_1,\dots,u_N)}.
 \end{split}
\end{equation}

 \item By the interpolation formulae we can solve the equation
       \eqref{sov-eq} and express the Hamiltonian $u_n$ in terms of the
       variables $(\lambda_j, z_j)_{j=1,\dots,N}$; $u_n =
       u_n(\lambda_1,\dots,\lambda_n; z_1,\dots,z_N)$. Thus we obtain a
       non-trivial integrable (or integrated) system.

\end{enumerate}

We have three canonical coordinate systems, $(\alpha_j,\psi_j)$, $(u_n,
\phi_n)$ and $(\lambda_j, \mu_j)$. The first two are of action-angle
type while the third are {\em separated variables}. In fact,
\eqref{sov-eq} may be interpreted as the equation of the 
common level set $u_n = u_n(z_1,\dots, z_N; \lambda_1,\dots,
\lambda_n)$ of the Hamiltonians expressed in 
separated variables (see \cite{bib:Sk}).  Note that, 
in general, if one solves the equations of the common level set 
by the implicit function theorem, the result would be
\begin{equation*}
    z_j = \Phi_j(\lambda_1,\dots,\lambda_N;u_1,\dots,u_N).
\end{equation*}
We emphasize that the right hand side of \eqref{sov-eq} contains only
$\lambda_j$ among all $\lambda_1,\dots,\lambda_N$. This is the key point
in the SoV method. 

Usually the SoV method requires a relation of the form
$\Psi_j(z_j,\lambda_j;u_1,\dots,u_N) = 0$ for each $j$. Our assumption
\eqref{sov-eq} says more than that; $N$ points $(\lambda_j, z_j)$ lie on
one and the same curve $C_u := \{(\lambda,z)\mid z =
A(\lambda;u_1,\dots,u_N)\}$ parametrized by $u_1,\dots,u_N$.

In this respect, the definition \eqref{sov-gen-func} of the generating
function of the canonical transformation is the Abelian
integral of the {\em Seiberg-Witten differential} $\log A(\lambda;u)
d\lambda$ on the curve $C_u$. We can also interpret the coordinate
transformation $(\alpha_1,\dots,\alpha_N) \mapsto (u_1,\dots,u_N)$ as
the inverse map of the {\em period integral},
\begin{equation}
    \alpha_j = \oint_{a_j} 
    \lambda\, d_\lambda \log A(\lambda; u_1,\dots,u_N),
\end{equation}
where $a_j$ is the cycle on the curve $C_u$ encircling a zero of
$A(\lambda;u)$ on the $\lambda$-plane. The interpolation formulae $u_n =
u_n(\lambda; z)$ determines the moduli of the curve $C_u$ from the
points lying on it.

\section{Rational case}
\label{sec:rat}

Let us consider the moduli space of rational functions
$B(\lambda)/A(\lambda)$ where $A$ and $B$ are polynomials of the form:
\begin{align}
    A(\lambda) 
    &= 
    \lambda^N + u_1 \lambda^{N-1} + u_2 \lambda^{N-2} + \cdots + u_N
    =
    \prod_{j=1}^N (\lambda-\alpha_j),
\label{A:rat}
\\
    B(\lambda)
    &=
    \rho \prod_{k=1}^{N-1} (\lambda-\lambda_k).
\label{B:rat}
\end{align}
For simplicity, we assume that all roots of $A$ and $B$ are
distinct. Since $B(\lambda)/A(\lambda)$ has a partial fraction
expansion,
\begin{equation}
    \frac{B(\lambda)}{A(\lambda)}
    =
    \sum_{j=1}^N \frac{\rho_j}{\lambda-\alpha_j},
    \qquad
    \rho=\sum_{j=1}^N \rho_j, 
    \qquad
    \rho_j = \frac{B(\alpha_j)}{A'(\alpha_j)},
\label{B/A:part-frac}
\end{equation}
there are two coordinates of this $2N$-dimensional moduli space,
$(\alpha_1,\dots,\alpha_N, \rho, \lambda_1,\dots, \lambda_{N-1})$ and
$(\rho_1, \dots, \rho_N, \alpha_1, \dots, \alpha_N)$.

Following Atiyah and Hitchin \cite{bib:At-Hi}, we introduce the
symplectic form
\begin{equation}
    \Omega = \sum_{j=1}^N d\log B(\alpha_j) \wedge d\alpha_j
\label{AH-form:rat}
\end{equation}
on this space. Since each $u_n$ is
the elementary symmetric function of $\alpha_j$, they commute with each
other; $\{u_m, u_n\} = 0$.

By an easy residue calculus, we can rewrite this form as follows:
\begin{equation}
    \Omega = \sum_{j=1}^N
    \Res_{\lambda=\alpha_j}\left(
    \frac{\delta A(\lambda)}{A(\lambda)} \wedge
    \frac{\delta B(\lambda)}{B(\lambda)} \wedge d\lambda
    \right),
\label{AH-form:res}
\end{equation}
where $\delta$ is the exterior derivation with respect to the moduli
coordinates, i.e., $\delta = d - d\lambda (\rd/\rd\lambda)$. The 3-form
inside the bracket is rational on the $\lambda$-sphere with poles at
$\lambda=\alpha_j$ ($j=1,\dots, N$), $\lambda=\lambda_k$
($k=1,\dots,N-1$) and $\lambda=\infty$. The residues at
$\lambda=\lambda_k$ and $\lambda=\infty$ are
\begin{align*}
    \Res_{\lambda=\lambda_k}\left(
    \frac{\delta A(\lambda)}{A(\lambda)} \wedge
    \frac{\delta B(\lambda)}{B(\lambda)} \wedge d\lambda
    \right) &= - d\log A(\lambda_k) \wedge d\lambda_k,
\\
    \Res_{\lambda=\infty}\left(
    \frac{\delta A(\lambda)}{A(\lambda)} \wedge
    \frac{\delta B(\lambda)}{B(\lambda)} \wedge d\lambda
    \right) &= - du_1 \wedge d\log\rho.    
\end{align*}
By the residue theorem, we have another expression of the form $\Omega$
from \eqref{AH-form:res},
\begin{equation}
    \Omega = \sum_{k=1}^{N-1} d\log A(\lambda_k) \wedge d\lambda_k
           + d u_1 \wedge d\log\rho.
\label{AH-form:A}
\end{equation}

Let us consider the restricted moduli space $\{ \rho=1, u_1 = 0\}$ (the
``reduced monopole space''; see \cite{bib:At-Hi}, p.19) and the
restriction of $\Omega$:
\begin{equation}
    \Omega|_{\rho=1,u_1=0} 
    = \sum_{k=1}^{N-1}  d\log A(\lambda_k) \wedge d\lambda_k.
\label{AH-form:restricted:A}
\end{equation}
Substituting the explicit expression \eqref{A:rat}, we have
\begin{equation}
 \begin{split}
    \Omega|_{\rho=1,u_1=0} 
    &=
    \sum_{k=1}^{N-1} \sum_{j=2}^N 
    \frac{\lambda_k^{N-j} du_j \wedge d\lambda_k}{A(\lambda_k)}
\\
    &=
    \sum_{j=2}^{N} du_j \wedge
    \left(
     \sum_{k=1}^{N-1} \frac{\lambda_k^{N-j}}{A(\lambda_k)} d\lambda_k
    \right).
 \end{split}
\label{AH-form:restricted:temp}
\end{equation}
We introduce the generating function of the canonical transformation by
\begin{equation}
    S(\lambda_1,\dots,\lambda_{N-1}; u_2,\dots,u_N)
    :=
    \sum_{k=1}^{N-1}
    \int^{\lambda_k}
    \log A(\lambda) \, d\lambda.
\label{gen-func:rat}
\end{equation}
The form $dS = \log A(\lambda)\, d\lambda$ is a Seiberg-Witten
differential on the curve $C_u = \{(\lambda,z)\mid z = A(\lambda)\}$ and
the right hand side of \eqref{gen-func:rat} is an Abelian
integral of $dS$ corresponding to the divisor $\sum_{k=1}^{N-1}
(\lambda_k, z_k)$ on $C_u$.

The variable $\mu_k = \log z_k$ and the ``angle variable'' $\phi_n$
corresponding to $u_n$ is expressed as
\begin{equation}
    \mu_k = \frac{\rd S}{\rd \lambda_k}, \qquad
    \phi_n = \frac{\rd S}{\rd u_n} 
    = \sum_{k=1}^{N-1} \int^{\lambda_k}
    \frac{\lambda^{N-j}}{A(\lambda)} d\lambda,
\label{phi:rat}
\end{equation}
and we can rewrite \eqref{AH-form:restricted:temp} as
\begin{equation}
    \Omega|_{\rho=1,u_1=0} 
    =
    \sum_{j=2} ^{N} du_j \wedge d\phi_n.
\label{AH-form:restricted:u,phi}
\end{equation}

The Hamiltonians $u_n$ ($n=2,\dots,N$) are explicitly written down in
terms of the coordinate system
$(\lambda_1,\dots,\lambda_{N-1},z_1,\dots,z_{N-1})$, where $z_k =
A(\lambda_k)$. This is a simple application of the Lagrange
interpolation formula,
\begin{equation}
    \frac{A(\lambda)-\lambda^N}{B(\lambda)}
    =
    \sum_{k=1}^{N-1}
    \frac{A(\lambda_k)-\lambda_k^N}{B'(\lambda_k)}
    \frac{1}{\lambda-\lambda_k}
    =
    \sum_{k=1}^{N-1}
    \frac{z_k-\lambda_k^N}{B'(\lambda_k)}
    \frac{1}{\lambda-\lambda_k}.
\label{interpolation:rat}
\end{equation}
(Note that the left hand side has poles at $\lambda=\lambda_k$ by
\eqref{B:rat}.) For example, $u_2$ is read off directly from the residue
of \eqref{interpolation:rat} at $\lambda=\infty$:
\begin{equation}
    u_2
    =
    \sum_{k=1}^{N-1} \frac{z_k-\lambda_k^N}{B'(\lambda_k)}
    =
    \sum_{k=1}^{N-1} 
    \frac{z_k-\lambda_k^N}{\prod_{l\neq k}(\lambda_k-\lambda_l)}.
\label{u2:rat}
\end{equation}
If we denote the coefficients of $B$ by $v_j$,
\begin{equation}
    B(\lambda) = \lambda^{N-1} + 
    \sum_{j=1}^N v_j \lambda^{N-j-1},
\label{B:expand:rat}
\end{equation}
other $u_j$'s can be expressed as
\begin{equation}
    u_{j+1}
    =
    - \sum_{k=1}^{N-1}
    \frac{z_k - \lambda_k^N}{B'(\lambda_k)}
    \frac{\rd v_j}{\rd \lambda_k},
\label{uj:rat}
\end{equation}
for $j=1,\dots,N-1$. (This expression recovers \eqref{u2:rat}, since
$v_1 = - \sum_{k=1}^{N-1} \lambda_k$.)

\section{Trigonometric (hyperbolic) case}
\label{sec:tri}

In the trigonometric case, we consider the moduli space of meromorphic
functions $B(\lambda)/A(\lambda)$ where $A$ and $B$ are trigonometric
polynomials:
\begin{align}
    A(\lambda) 
    &= 
    \prod_{j=1}^N \sh (\lambda-\alpha_j),
\label{A:tri}
\\
    B(\lambda)
    &=
    \prod_{k=1}^{N} \sh(\lambda-\lambda_k).
\label{B:tri}
\end{align}
For simplicity, 
we assume that $\lambda_k$'s and $\alpha_j$'s are distinct. Since
$B(\lambda)/A(\lambda)$ has a period $\pi i$, $\alpha_j$ and $\lambda_k$
are regarded as points on the cylinder, $\bbC/\pi i \bbZ$.

The symplectic form
\begin{equation}
    \Omega = \sum_{j=1}^N d\log B(\alpha_j) \wedge d\alpha_j
\label{AH-form:tri}
\end{equation}
is a trigonometric analogue of the form \eqref{AH-form:rat}. The
expression \eqref{AH-form:res} holds also in this case. To obtain the
formula like \eqref{AH-form:A} or \eqref{AH-form:restricted:A}, we can
apply the residue formula which expresses the sum of the residues as the
contour integral, but the following argument is simpler: the logarithmic
derivative of $\sh \lambda$ is the hyperbolic cotangent, $\cth \lambda$,
which is an odd function, $\cth(-\lambda) = - \cth\lambda$. Therefore it
follows from the definition \eqref{AH-form:tri} that
\begin{equation}
 \begin{split}
    \Omega &=
    -\sum_{j,k=1}^N \cth(\alpha_j-\lambda_k) d\lambda_k \wedge d\alpha_j
\\
    &=
    -\sum_{j,k=1}^N \cth(\lambda_k-\alpha_j) d\alpha_j \wedge d\lambda_k
\\
    &= \sum_{k=1}^N d\log A(\lambda_k) \wedge d\lambda_k.
 \end{split}
\label{AH-form:tri:A}
\end{equation}

As in the rational case the coefficients of $A(\lambda)$ become action
variables. Let us denote $e^\lambda$ by $x$ and expand $A(\lambda)$ as
\begin{equation}
    A(\lambda) =
    \frac{1}{2^N}(
    u_0 x^{N} - u_1 x^{N-2} + \cdots 
    + (-1)^{N-1} u_{N-1} x^{2-N} + (-1)^N u_N x^{-N}
    ).
\label{A:tri:expand}
\end{equation}
Each coefficient $u_n$ is expressed explicitly as
\begin{equation}
    u_n = \sum_{\substack{I_1 \sqcup I_2 = \{1,\dots,N\}\\ |I_1| = n}}
    \prod_{i\in I_1} e^{\alpha_i} \prod_{j\in I_2} e^{-\alpha_j}.
\label{un:tri}
\end{equation}
In particular,
\begin{equation}
    u_0 = \exp\left(-\sum_{j=1}^N \alpha_j\right), \qquad
    u_N = u_0^{-1}.
\label{u0,uN:tri}
\end{equation}
Since $u_n$'s are functions of $\alpha_j$, they commute with each other
with respect to the Poisson bracket defined by the form $\Omega$,
\eqref{AH-form:tri}. 

Taking the formula \eqref{u0,uN:tri} into account, we have
\begin{equation}
    d \log A(\lambda_k) \wedge d\lambda_k
    =
    \frac{1}{2^N}
    \left(
    \sum_{n=1}^{N-1} 
    du_n \wedge \frac{(-1)^n x_k^{N-2n}}{A(\lambda_k)} d\lambda_k
    +
    \frac{du_0}{u_0} \wedge 
    \frac{u_0 x_k^N - (-1)^N u_0^{-1} x_k^{-N}}{A(\lambda_k)}d\lambda_k
    \right),
\label{AH-form:tri:u,phi:temp}
\end{equation}
where $x_k = e^{\lambda_k}$. Summing them up, we have from
\eqref{AH-form:tri:A} that
\begin{equation}
    \Omega = \sum_{n=1}^{N-1} du_n \wedge d\phi_n
    + \frac{du_0}{u_0} \wedge d\phi_0,
\label{AH-form:tri:u,phi}
\end{equation}
where $\phi_n$'s are angle variables defined by
\begin{equation}
    \phi_n = \frac{1}{2^N}\sum_{k=1}^N
    \int^{\lambda_k} 
    \frac{(-1)^n e^{(N-2n)\lambda}}{A(\lambda)} 
    d\lambda,
\label{phi:tri}
\end{equation}
for $n=1,\dots,N-1$ and
\begin{equation}
    \phi_0 = \frac{1}{2^N}\sum_{k=1}^N
    \int^{\lambda_k} 
    \frac{u_0 e^{N\lambda} - (-1)^N u_0^{-1} e^{-N\lambda}}{A(\lambda)} 
    d\lambda.
\label{phi0:tri}
\end{equation}
The generating function of the canonical transformation $(\lambda_k,
\mu_k = \log A(\lambda_k))_{k=1,\dots,N} \mapsto
(u_n,\phi_n)_{n=1,\dots,N}$ is
\begin{equation}
    S(\lambda_1,\dots,\lambda_N; u_1,\dots,u_N)
    =
    \sum_{k=1}^{N} \int^{\lambda_k}
    \log A(\lambda)\, d\lambda,
\label{gen-func:tri}
\end{equation}
and the variable $\mu_k$ and $\phi_n$ are expressed as
\begin{equation}
    \mu_k = \frac{\rd S}{\rd \lambda_k}, \qquad
    \phi_n = \frac{\rd S}{\rd u_n}.
\end{equation}
The right hand side of \eqref{gen-func:tri} is again the Abelian
integral of the Seiberg-Witten differential $\log A(\lambda)\, d\lambda$
on the curve $C_u = \{(\lambda,z)\mid z = A(\lambda)\}$ parametrized by
$u_1,\dots,u_N$.

Explicit expressions of the Hamiltonians $u_n$ ($n=0,\dots,N-1$) in
terms of the coordinate system
$(\lambda_1,\dots,\lambda_{N-1},z_1,\dots,z_{N-1})$, $z_k =
A(\lambda_k)$, are obtained from the interpolation formula. By comparing
the periodicity, position of poles and their residue, it is easy to see
that
\begin{equation}
    \frac{A(\lambda)}{B(\lambda)}
    =
    \sum_{k=1}^N \frac{z_k}{B'(\lambda_k)}\cth(\lambda-\lambda_k)
    + c_0,
\label{interpolation:tri}
\end{equation}
where $c_0$ is a constant independent of $\lambda$ but dependent on
$\lambda_k$ and $\alpha_j$. Let us expand $B(\lambda)$ as
\begin{equation}
    B(\lambda) = 
    \frac{1}{2^N}(
    v_0 x^{N} - v_1 x^{N-2} + \cdots 
    + (-1)^{N-1} v_{N-1} x^{2-N} + (-1)^N v_N x^{-N}
    ),
\label{B:tri:expand}
\end{equation}
where
\begin{equation}
    v_n = \sum_{\substack{I_1 \sqcup I_2 = \{1,\dots,N\}\\ |I_1| = n}}
    \prod_{i\in I_1} e^{\lambda_i} \prod_{j\in I_2} e^{-\lambda_j}.
\label{vn:tri}
\end{equation}
In particular, $v_0 = \exp\left(- \sum_{k=1}^N \lambda_k\right)$, $v_N =
v_0^{-1}$. Summing up the asymptotics of the both hand sides of
\eqref{interpolation:tri} in the limit $\Re \lambda \to \pm\infty$,
we have 
\begin{equation}
   \frac{u_0}{v_0} + \frac{v_0}{u_0} = 2 c_0 ,
   \text{ i.e.,\ } 
   c_0
   = \frac{1}{2}\left(\frac{u_0}{v_0} + \frac{v_0}{u_0}\right)
   = \ch\left(\sum_{k=1}^N \lambda_k - \sum_{j=1}^N \alpha_j\right),
\label{const:tri}
\end{equation}
because $\cth\lambda \to \pm 1$ ($\Re\lambda \to \pm\infty$). 
Expansion of \eqref{interpolation:tri} around $x=e^\lambda =\infty$
gives
\begin{equation}
\begin{split}
    &u_0 - u_1 x^{-2} + \cdots 
    + (-1)^{N-1} u_{N-1} x^{2-2N} + (-1)^N u_N x^{-2N}
    \\ = &
    (v_0 - v_1 x^{-2} + \cdots 
    + (-1)^{N-1} v_{N-1} x^{2-2N} + (-1)^N v_N x^{-2N})
    \\ \times &
    \left(c_0 + 
    \sum_{k=1}^N 
    (1 + 2x^{-2}e^{2\lambda_k}+ 2x^{-4}e^{4\lambda_k}+\cdots)
    \frac{z_k}{B'(\lambda_k)}\right).
\end{split}
\label{recursion:tri}
\end{equation}
(Note that $\cth \lambda = 1 + \sum_{n=1}^\infty 2 x^{-2n}$ for $|x| >
1$.) By comparing the coefficients of $x^{-2n}$ we obtain the expression
of $u_n$. For example, the coefficients of $x^0$ is
\begin{equation}
    u_0 
    = 
    v_0 \left( c_0 + \sum_{k=1}^N \frac{z_k}{B'(\lambda_k)}\right),
\label{u0:tri}
\end{equation}
which is equivalent to
\begin{equation}
    \sum_{k=1}^N \frac{z_k}{B'(\lambda_k)}
    = \frac{1}{2} \left(\frac{u_0}{v_0} - \frac{v_0}{u_0}\right)
    = \sh\left(\sum_{j=1}^N \lambda_j - \sum_{k=1}^N \alpha_j\right)
\label{res-sum:tri}
\end{equation}
due to \eqref{const:tri}\footnote{We can derive this relation directly
from the interpolation formula \eqref{interpolation:tri}. In fact, the
left hand side of \eqref{res-sum:tri} is the sum of residues of
\eqref{interpolation:tri}. The right hand side is the result of a
contour integral of $A(\lambda)/B(\lambda)$ along the rectangle with
vertices $R$, $R + \pi i$, $-R + \pi i$, $-R$. Since $\lim_{\Re \lambda
\to \pm\infty} A(\lambda)/B(\lambda) = \prod_{j=1}^N \exp(\pm(\lambda_j
- \alpha_j))$, this contour integral becomes $2\pi i$ times the right
hand side of \eqref{res-sum:tri}.}. Therefore $u_0$ is the solution of
the quadratic equation
\begin{equation}
    u_0^2 
    - 2 \left(\sum_{k=1}^N \frac{z_k}{B'(\lambda_k)}\right) v_0 u_0
    - v_0^2 = 0.
\end{equation}

The coefficients of $x^{-2}$ of \eqref{recursion:tri} gives the
expression of $u_1$ as follows:
\begin{equation}
    u_1 = 2 v_0 \sum_{k=1}^N \frac{e^{2\lambda_k} z_k}{B'(\lambda_k)}
        - \frac{v_1}{v_0} u_0,
\label{u1:tri}
\end{equation}
and other $u_n$'s are determined recursively.

\section{Elliptic case}
\label{sec:ell}

In the elliptic case, we take the following elliptic polynomials
$A(\lambda)$ and $B(\lambda)$: 
\begin{align}
    A(\lambda) &= \prod_{j=1}^N \sigma(\lambda-\alpha_j),
\label{A:ell}
\\
    B(\lambda) &= \prod_{k=1}^N \sigma(\lambda-\lambda_k),
\label{B:ell}
\end{align}
where $\sigma(\lambda)=\sigma(\lambda;2\omega_1,2\omega_3)$ is the
Weierstrass $\sigma$ function with periods $2\omega_1$ and
$2\omega_3$.
The symplectic form
\begin{equation}
    \Omega = \sum_{j=1}^N d\log B(\alpha_j) \wedge d\alpha_j
\label{AH-form:ell}
\end{equation}
is defined as in the rational and the trigonometric cases. Due to the
oddness of the Weierstrass $\zeta$ function $\zeta(-\lambda) =
-\zeta(\lambda)$, we have
\begin{equation}
 \begin{split}
    \Omega &=
    -\sum_{j,k=1}^N \zeta(\alpha_j-\lambda_k) d\lambda_k \wedge d\alpha_j
\\
    &=
    -\sum_{j,k=1}^N \zeta(\lambda_k-\alpha_j) d\alpha_j \wedge d\lambda_k
\\
    &= \sum_{k=1}^N d\log A(\lambda_k) \wedge d\lambda_k.
 \end{split}
\label{AH-form:ell:A}
\end{equation}

We set $u_0 = \sum_{j=1}^N \alpha_j$ and regard $u_0$ and the
coefficients in expansion of $A(\lambda)$,
\begin{equation}
    A(\lambda) = \sum_{n=1}^{N} u_n f_n(\lambda;u_0),
\label{A:ell:expand}
\end{equation}
as commuting Hamiltonians, where each $f_n(\lambda;u_0)$ has the same
quasi-periodicity with $A(\lambda)$, namely,
\begin{equation}
    f_n(\lambda+2\omega_i;u_0) 
    = (-1)^N e^{2\eta_i(N\lambda - u_0)} f_n(\lambda;u_0), \qquad
    \eta_i := \zeta(\omega_i),
\label{quasi-period}
\end{equation}
for $i=1,3$. (The space of functions with quasi-periodicity
\eqref{quasi-period} is $N$-dimensional.) For example, the functions
\begin{equation}
    f_n(\lambda;u_0) = \sigma(\lambda)^N \phi_{u_0}^{(N-n)}(\lambda), 
    \qquad
     n=1,\dots,N,
\label{fn}
\end{equation}
fit our purpose, where $\phi_{c}(\lambda) =
\sigma(\lambda-c)/\sigma(\lambda)\sigma(-c)$, but in the following we do
not need the explicit form.

The coefficients $u_0,\dots, u_N$ in the expansion \eqref{A:ell:expand}
are functions of the variables $\alpha_j$'s and do not depend on
$B(\lambda)$. Hence they commute with each other with respect to the
Poisson bracket; $\{u_m, u_n\} = 0$.

But they {\em cannot} be independent because $A(\lambda)$ depends on $N$
parameters $\{\alpha_j\}$.  Regarding, for example, $u_N$ as a function
of independent parameters $u_0,\dots,u_{N-1}$, we rewrite the symplectic
form $\Omega$ as follows:
\begin{equation}
 \begin{split}
    \Omega &=
    \sum_{k=1}^{N} d\log A(\lambda_k) \wedge d\lambda_k
    =
    \sum_{k=1}^{N} 
    \frac{d A(\lambda_k)}{A(\lambda_k)} \wedge d\lambda_k
\\
    &=
    \sum_{k=1}^{N} \sum_{n=0}^{N-1}
    du_n \wedge 
    \frac{\rd A(\lambda_k)}{\rd u_n}
    \frac{d\lambda_k}{A(\lambda_k)} 
\\
    &=
    \sum_{k=1}^{N}
    du_0 \wedge 
    \left(
    \sum_{l=1}^{N} u_l \frac{\rd f_l}{\rd u_0}(\lambda_k;u_0)
    + \frac{\rd u_N}{\rd u_0} f_N(\lambda_k;u_0)\right)
    \frac{d\lambda_k}{A(\lambda_k)}
\\
    &+ 
    \sum_{k=1}^{N}
    \sum_{n=1}^{N-1}
    du_n \wedge 
    \left(
    f_n(\lambda_k;u_0)+
    \frac{\rd u_N}{\rd u_n} f_N(\lambda_k;u_0)
    \right)
    \frac{d\lambda_k}{A(\lambda_k)}.
 \end{split}
\label{AH-form:ell:u,phi:temp}
\end{equation}
Hence, introducing the generating function of the canonical
transformation as the Abelian integral of the Seiberg-Witten
differential $\log A(\lambda)\, d\lambda$ on the curve $C_u =
\{(\lambda,z) \mid z = A(\lambda)\}$ (a subvariety of the total space
of the line bundle over the elliptic curve defined by the
quasi-periodicity \eqref{quasi-period}),
\begin{equation}
    S(\lambda_0,\dots, \lambda_{N-1}; u_0,\dots,u_{N-1})
    =
    \sum_{k=0}^N \int^{\lambda_k}
    \log A(\lambda)\, d\lambda,
\label{gen-func:ell}
\end{equation}
we can express the ``angle variable'' $\phi_n$ by
\begin{equation}
 \begin{split}
    \phi_0 &= \frac{\rd S}{\rd u_0}
    =
    \sum_{k=1}^{N} \int^{\lambda_k}
    \left(
    \sum_{l=1}^{N} u_l \frac{\rd f_l}{\rd u_0}(\lambda;u_0)
    + \frac{\rd u_N}{\rd u_0} f_N(\lambda;u_0)\right)
    \frac{d\lambda}{A(\lambda)},
\\
    \phi_n &= \frac{\rd S}{\rd u_n}
    = \sum_{k=1}^{N} \int^{\lambda_k}
    \left(f_n(\lambda;u_0)
          + \frac{\rd u_N}{\rd u_n} f_N(\lambda;u_0)\right)
    \frac{d\lambda}{A(\lambda)}
 \end{split}
\label{phi:ell}
\end{equation}
for $n=1,\dots,N-1$. The canonical $2$-form $\Omega$ is expressed as
\begin{equation}
    \Omega= \sum_{n=0}^{N-1} du_n \wedge d\phi_n.
\label{AH-form:ell:u,phi}
\end{equation}

Hamiltonians $u_n$ have similar expressions as the trigonometric case in
terms of the coordinates $\lambda_k$, $z_k = A(\lambda_k)$. There are
$N$ equations to be satisfied:
\begin{equation}
    z_j = \sum_{n=1}^{N} u_n f_n(\lambda_j;u_0)
\label{interpolation:ell:eq}
\end{equation}
for $j=1,\dots,N$, or equivalently,
\begin{equation}
    \begin{pmatrix} z_1 \\ z_2 \\ \vdots \\ z_N \end{pmatrix}
    =
    \begin{pmatrix}
    f_1(\lambda_1;u_0) &f_2(\lambda_1;u_0) &\dots &f_{N}(\lambda_1;u_0) \\
    f_1(\lambda_2;u_0) &f_2(\lambda_2;u_0) &\dots &f_{N}(\lambda_2;u_0) \\
    \vdots & \vdots & \ddots & \vdots \\
    f_1(\lambda_N;u_0) &f_2(\lambda_N;u_0) &\dots &f_{N}(\lambda_N;u_0) 
    \end{pmatrix}
    \begin{pmatrix} u_1 \\ u_2 \\ \vdots \\ u_N \end{pmatrix}.
\label{interpolation:ell:eq:mat}
\end{equation}
The determinant of the matrix in the right hand side factors as follows:
\begin{equation}
 \begin{split}
    D(\lambda_1,\dots,\lambda_N;u_0) &:=
    \left|\begin{matrix}
    f_1(\lambda_1;u_0) &f_2(\lambda_1;u_0) &\dots &f_{N}(\lambda_1;u_0) \\
    f_1(\lambda_2;u_0) &f_2(\lambda_2;u_0) &\dots &f_{N}(\lambda_2;u_0) \\
    \vdots & \vdots & \ddots & \vdots \\
    f_1(\lambda_N;u_0) &f_2(\lambda_N;u_0) &\dots &f_{N}(\lambda_N;u_0) 
    \end{matrix}\right|
\\
    &= 
    (\text{non-zero constant})\times
    \sigma\left(\sum_{i=1}^{N}\lambda_i - u_0 \right)
    \prod_{1\leqq i < j \leqq N} \sigma(\lambda_i - \lambda_j),
 \end{split}
\label{det}
\end{equation}
The proof is given in Appendix \ref{app:det}. Therefore the equation
\eqref{interpolation:ell:eq:mat} is generically non-degenerate and
solved as:
\begin{equation}
 \begin{split}
    A(\lambda) 
    =& u_1 f_1(\lambda;u_0) + u_2 f_2(\lambda;u_0) + \cdots 
     + u_{N-1} f_{N-1}(\lambda;u_0)
\\
    =&
    D(\lambda_1,\dots,\lambda_N;u_0)^{-1}
    \left|\begin{matrix}
    0 &     f_1(\lambda;u_0)   & \dots  & f_N(\lambda;u_0) \\
    - z_1 & f_1(\lambda_1;u_0) &  \dots & f_N(\lambda_1;u_0) \\
    \vdots & \vdots & \ddots & \vdots \\
    - z_N & f_1(\lambda_N;u_0) &  \dots & f_N(\lambda_N;u_0) 
    \end{matrix}\right|.
 \end{split}
\label{interpolation:ell}
\end{equation}

In this formula $u_0$ has not yet been expressed as the function of
$\lambda_k$ and $z_k$. It is determined as an implicit function. Taking
the logarithm of \eqref{A:ell}, we have
\begin{equation*}
 \begin{split}
    \log A(\mu) &=
    \sum_{j=1}^N \log \sigma(\mu - \alpha_j)
    =
    \sum_{\lambda=\text{root of $A(\lambda)$}} \log \sigma(\mu-\lambda)
\\
    &=
    \frac{1}{2\pi i} \oint
    \log \sigma(\mu - \lambda) d_\lambda \log A(\lambda),
 \end{split}
\end{equation*}
where the integration contour surrounds each $\alpha_j$ once but not
$\mu + 2\omega_1 n_1 + 2\omega_3 n_3$ ($n_1, n_3 \in \bbZ$). Hence,
e.g., fixing $\mu$ to $0$, we have an equation
\begin{equation}
    \log A(0) =
    \frac{1}{2\pi i} \oint
    \log \sigma(-\lambda) d_\lambda \log A(\lambda).
\label{logA}
\end{equation}
Substituting \eqref{interpolation:ell} into this equation, we have an
equation which fixes $u_0$.

\section{Perspective from symplectic surfaces}
\label{sec:sur}

Let us reconsider the integrable systems of the three 
types from the point of view of symplectic or Poisson 
surfaces \cite{bib:Hu,bib:Va,bib:Bo}.  

The complex surface $X$ for the rational case 
is essentially the $(z,\lambda)$ plane with 
the line $z = 0$ deleted.  The symplectic structure 
is defined by the 2-form 
\begin{equation}
    \omega = \frac{dz \wedge d\lambda}{z}. 
\end{equation}
This surface is the affine part of a rational 
surface fibered over $\bbP^1$.  $dz/z$ is 
a holomorphic differential on the fibers 
$\bbC^* = \bbC \setminus \{0\}$ (i.e., cylinders).  
The symplectic structure on $X$ induces a symplectic 
structure on the (smooth part of) ($N-1$)-fold 
symmetric product $X^{(N-1)} = X^{N-1}/S_{N-1}$ 
with the symplectic form 
\begin{equation}
    \Omega 
  = \sum_{j=1}^{N-1} \frac{dz_j \wedge d\lambda_j}{z_j}, 
\end{equation}
where $(z_j,\lambda_j)_{j=1}^{N-1}$ is an $(N-1)$-tuple 
of points of $X$ that represent a point of $X^{(N-1)}$.  
The interpolation formulae \eqref{u2:rat} and 
\eqref{uj:rat} imply that an $(N-1)$-tuple 
of points of $X$ in {\em general position} 
uniquely determines a curve of the form $C_u = 
\{(\lambda,z) \mid z = A(\lambda)\}$ that passes 
the $N-1$ points.  We thus obtain a mapping 
\begin{equation}
    (\lambda_j,z_j)_{j=1}^{N-1} \longmapsto 
    (u_2,\ldots,u_N) 
\end{equation}
from an open subset of $X^{(N-1)}$ to $\bbC^{N-1}$, 
the fibers of which are Lagrangian subvarieties with 
respect to $\Omega$.   This is a geometric interpretation 
of the integrable system on the moduli spaces of rational  
functions.  

The same interpretation carries over to the trigonometric and 
elliptic cases.  The symplectic surface $X$ for these cases 
is also cylindrically fibered over a Riemann surface 
$\Sigma$ (cylinder or torus), and the symplectic form 
can be written in the same form 
\begin{equation*}
    \omega = \frac{dz \wedge d\lambda}{z} 
\end{equation*}
except that $\lambda$ is a coordinate on $\Sigma$.  

We now turn to integrable systems associated with 
an elliptically fibered symplectic surface \cite{bib:Ta-K3}.   
It should be noted that some part of the structure 
outlined in Section \ref{sec:out} is no longer retained 
or largely modified.  In particular, there is no counterpart 
of the first system of canonical coordinates $(\alpha_j,\psi_j)$; 
it is the second and third systems of canonical coordinates 
(i.e., action-angle variables and ``separated variables'') 
that play a central role.  

The first example is a specialization of Beauville's 
integrable systems \cite{bib:Be} to a K3 surface $X$ 
with elliptic fibration $X \to \bbP^1$.   An affine 
model of this surface is defined by the equation 
\begin{equation}
    y^2 = 4z^3 + g_2(\lambda) z + g_3(\lambda) 
    \label{ell-surface}
\end{equation}
of the Weierstrass normal form.  $g_2(\lambda)$ and 
$g_3(\lambda)$ are (generic) polynomials of degrees 
$8$ and $12$, respectively.  The complex symplectic 
structure is defined by the 2-form 
\begin{equation}
    \omega = \frac{dz \wedge d\lambda}{y}. 
    \label{ell-omega}
\end{equation}
Note that the differential $dz/z$ along the cylindrical 
fibers is now replaced by $dz/y$ on the elliptic fibers. 
A canonically conjugate variable of $\lambda$ is 
given by the elliptic integral (along the fibers 
of $X \to \bbP^1$) 
\begin{equation}
    \mu(z,\lambda) 
    = \int_{(\infty,\infty)}^{(\lambda,z)}\frac{dz}{y}, 
\end{equation}
the inversion of which is given by the Weierstrass $\wp$ 
function $z = \wp(\mu)$ with $\lambda$-dependent 
primitive periods.  

The construction of the integrable system proceeds 
as follows \cite{bib:Ta-K3}: 
\begin{enumerate}
\item 
Choose a five-parameter family of curves $C_u$ in $X$ 
cut out by the equation 
\begin{equation}
    z = A(\lambda) = \sum_{n=1}^5 u_n\lambda^{5-n}. 
\end{equation}
$C_u$ is thus a hyperelliptic curve of genus $5$ 
defined by the equation 
\begin{equation}
    y^2 = 4A(\lambda)^3 + g_2(\lambda)A(\lambda) + g_3(\lambda). 
\end{equation}
\item 
The phase space of the integrable system is (an open 
subset of) the five-fold symmetric product $X^{(5)}$ 
equipped with the symplectic form 
\begin{equation}
    \Omega = \sum_{j=1}^5 \frac{dz_j \wedge d\lambda_j}{y_j}, 
\end{equation}
where the $5$-tuple $(\lambda_j,y_j,z_j)_{j=1}^5$ of 
points of $X$ represents a point of $X^{(5)}$.  
A set of canonical coordinates are given by 
$\lambda_j$ and $\mu_j = \mu(A(\lambda_j),\lambda_j)$, 
$j = 1,\ldots,5$.  If $\lambda_j$ are mutually distinct, 
the equations 
\begin{equation}
    z_j = A(\lambda_j) 
\end{equation}
can be solved for $u_j$, which are thus defined as 
functions on an open subset of $X^{(5)}$.  
\item 
Plugging the equations $z_j = A(\lambda_j)$ into 
the expression of $\Omega$ and doing some algebra, 
one finds that 
\begin{equation}
    \Omega = \sum_{n=1}^5 du_n \wedge d\phi_n, 
\end{equation}
where $\phi_n$ are defined by the Abel-Jacobi integrals 
\begin{equation}
    \phi_n 
  = \sum_{j=1}^5 \int_{(\infty,\infty)}^{(\lambda_j,y_j)}
    \frac{\lambda^{5-n}d\lambda}{y}, 
\end{equation}
of the the holomorphic differentials 
$\lambda^{5-n}d\lambda/y$ on $C_u$.  This expression 
of $\Omega$ shows that $u_n$ and $\phi_n$ are 
action-angle variables.  In particular, the fibers 
of the mapping $(\lambda_j,y_j,z_j)_{j=1}^5 \mapsto 
(u_1,\ldots,u_5)$ on an open subset of $X^{(5)}$ 
turns out to be Lagrangian subvarieties.  
\end{enumerate}
The ``separated variables'' $(\lambda_j,\mu_j)$ 
and the action-angle variables $(u_n,\phi_n)$ 
are connected by a canonical transformations. 
The generating function takes the form 
\begin{equation}
    S = \sum_{j=1}^5 \int_\infty^{\lambda_j} 
        \mu(A(\lambda),\lambda)d\lambda 
\end{equation}
and the canonical transformation is defined by 
\begin{equation}
    \frac{\rd S}{\rd \lambda_j} = \mu_j, \quad 
    \frac{\rd S}{\rd u_n} = \phi_n. 
\end{equation}
Behind this construction is the Seiberg-Witten differential 
\begin{equation}
    dS = \mu(A(\lambda),\lambda)d\lambda. 
\end{equation}
The logarithmic factor $\log A(\lambda)$ in the 
cylindrically fibered case is thus replaced by 
the elliptic integral $\mu(A(\lambda),\lambda)$.  

A variant of this integrable system is obtained from 
a rational surface with elliptic fibration \cite{bib:Ta-K3}.  
Such a surface, too, can be defined by the Weierstrass 
normal form \eqref{ell-surface}; $g_2(\lambda)$ and 
$g_3(\lambda)$ in this case are polynomials of degree 
$4$ and $6$.  The symplectic form $\omega$ \eqref{ell-omega} 
has poles along a compactification divisor at infinity.  
As a family of curves, we choose the two-parameter family 
$C_u$ cut out by the equation 
\begin{equation}
    z = A(\lambda) = c\lambda^2 + u_1\lambda + u_2.  
\end{equation}
$C_u$ is a hyperelliptic curve of genus $2$.  Note 
that the setting is slightly different from the case 
of the elliptically fibered K3 surface -- whereas $u_1$ 
and $u_2$ are moduli, $c$ should be treated as a central 
element (Casimir function) of a Poisson structure.  
Apart from this difference, the construction of 
an integrable system is fully parallel: 
The phase space is realized as (an open subset of) 
the two-fold symmetric product $X^{(2)}$; 
the Hamiltonians $u_1$ and $u_2$ are defined on 
an open subset of $X^{(2)}$ by the equations 
$z_j = A(\lambda_j)$; angle variables conjugate 
to $(u_1,u_2)$ are given by the Abel-Jacobi integrals 
\begin{equation}
  \phi_1 = 
  \sum_{j=1,2}\int_{(\infty,\infty)}^{(\lambda_j,y_j)}
  \frac{\lambda d\lambda}{y}, 
  \quad 
  \phi_2 = 
  \sum_{j=1,2}\int_{(\infty,\infty)}^{(\lambda_j,y_j)}\frac{d\lambda}{y}
\end{equation}
of holomorphic differentials on $C_u$; a generating 
function $S$ for these action-angle variables can be 
defined in exactly the same way.  

These two examples can be thought of as a generalization 
of the integrable system on the moduli space of rational 
functions.  In particular, the Hamiltonians $u_n$ are 
constructed by the same Lagrange interpolation formula, 
so that they take the same form \eqref{u2:rat} and 
\eqref{uj:rat}, once written in the (local) coordinates 
$(\lambda_k,z_k)$ on the symmetric product of $X$.  
One will be further tempted to increase the degrees 
of $A(\lambda)$, $g_2(\lambda)$ and  $g_3(\lambda)$ to, 
say, $2m$, $4m$ and $6m$ ($m = 2,3,\ldots$).  This 
is somewhat problematical:  The genus $3m - 1$ of 
the (still hyperelliptic) curve $C_u$ then exceeds 
the number $2m + 1$ of the moduli of the polynomial 
$A(\lambda)$.  One thus has to construct an 
integrable system from at most $(2m + 1)$-tuple 
of points on a family of curves of genus greater 
$2m + 1$ --- a considerably unusual setting.  
Nevertheless the construction seems to work at least 
formally.  

One can conversely start from the systems associated 
with an elliptically fibered surface, and consider 
the system on the moduli space of rational functions 
as a kind of degeneration.  From this point of view, 
we find some other types of degeneration in accordance 
with degeneration of the elliptic function $z = \wp(\mu)$ 
(and the associated singular rational curve), e.g., 
\begin{enumerate}
\item trigonometric (hyperbolic) function 
\begin{equation}
    \mu = \int^z \frac{dz}{2\sqrt{z}(z-1)}, \quad 
    z = \coth^2 \mu, \quad 
\end{equation}
\item quadratic function 
\begin{equation}
    \mu = \int^z \frac{dz}{2\sqrt{z}}, \quad 
    z = \mu^2, 
\end{equation}
\item exponential function 
\begin{equation}
    \mu = \int^z \frac{dz}{z}, \quad 
    z = e^{\mu},
\end{equation}
\item linear function 
\begin{equation}
    \mu = \int^z dz, \quad 
    z = \mu, 
\end{equation}
\end{enumerate}
which are also known to emerge in a correspondence 
between the Painlev\'e equations and the (generalized) 
Calogero systems \cite{bib:vD,bib:Ta-pcc}. The third one 
in this list is nothing but the case of the moduli space 
of rational functions.  The construction of an integrable 
system for the other cases is fully parallel to that case, 
except that the canonical coordinates $(\alpha_j,\psi_j)$ 
are no longer given by the zeros of $A(\lambda)$ etc.  

We can further seek generalization to elliptic fibration 
over a Riemann surface $\Sigma$ other than the sphere.  
Of particular interest is the case of an elliptically 
fibered surface over an elliptic curve.  Recent work of 
Braden et al. \cite{bib:Br-Go-Od-Ru} appears to provide 
a lot of material on this issue.

\section{Conclusion}
\label{sec:conc}

Starting from an integrable Hamiltonian system on the moduli
space of rational functions, we constructed several
integrable systems. As we have seen in \S\ref{sec:sur}, the
phase spaces of them are the symmetric product $X^{(N)}$ of
a complex symplectic surface $X$. The surface for the
systems considered in \S\S\ref{sec:rat}--\ref{sec:ell} are
the fibration over Riemann surfaces (the Riemann sphere, a
cylinder and a torus) with the cylinder as a fiber, while
the systems reviewed in \S\ref{sec:sur} are based on the
symplectic surfaces elliptically fibered over $\bbP^1$.

These systems have other common features besides their
algebro-geometric nature.
\begin{itemize}
 \item They have two specific sets of canonical variables:
       the action-angle variables $(u_n, \phi_n)$ and the
       separated variables $(\lambda_k, z_k)$. 

 \item The generating function of the canonical
       transformation between $(u_n, \phi_n)$ and
       $(\lambda_k, z_k)$ is the Abel-Jacobi integral of the
       Seiberg-Witten differential on the spectral curve.

 \item The Hamiltonians $(u_1,\dots,u_N)$ are explicitly
       described by the interpolation formula in terms of
       the separated variables $(\lambda_k, z_k)$. The map
       $(\lambda_k, z_k) \mapsto (u_1,\dots,u_N)$ gives a
       Lagrangian fibration of the phase space $X^{(N)}$. 
\end{itemize}

Our construction is quite explicit and easy to
generalize. One would have many other variants, using other
symplectic surfaces, but the explicit description of the
system would be more difficult.

\subsection*{Acknowledgements}

The authors are grateful to Harry Braden, Takeshi Ikeda, Yoshimasa
Nakamura and Orlando Ragnisco for their interests and comments.  The
authors are partly supported by the Grant-in-Aid for Scientific Research
(No.\ 12640169 and No.\ 13740005), Japan Society for the Promotion of
Science.

\appendix
\section{Proof of \eqref{det}}
\label{app:det}

In this appendix, we prove the formula \eqref{det}.

Let us denote the space of meromorphic functions $f(\lambda)$ of
$\lambda\in\bbC$ with the following properties by $\calL$:
\begin{itemize}
 \item the poles of which are located only on the lattice $2\omega_1 \bbZ 
       + 2 \omega_3 \bbZ$ and of order not greater than $N$.
 \item $f(\lambda+2\omega_i) = e^{-2\eta_i u_0} f(\lambda)$ for $i=1,3$.
\end{itemize}
The linear space $\calL$ is spanned by the functions,
\begin{equation}
    \phi(\lambda), \phi'(\lambda), \dots, \phi^{(N-1)}(\lambda),
\label{basis:A/sigma}
\end{equation}
where $\phi(\lambda) = \phi_{u_0}(\lambda)$. (See \eqref{fn}.)
Assume that $\sum_{i=0}^{N-1} \lambda_i = u_0$ and set $f(\lambda) =
\sigma(\lambda)^{-N} \prod_{i=0}^{N-1} \sigma(\lambda-\lambda_i)$, which
belongs to $\calL$. Expanding $f(\lambda)$ by the basis
\eqref{basis:A/sigma}, we have a system of $N$ linear equations,
\begin{equation}
    0 = a_{N-1} \phi(\lambda_j) + a_{N-2} \phi'(\lambda_j) + \cdots 
       + a_0 \phi^{(N-1)}(\lambda_j)
\label{f(lambda-j)=0}
\end{equation}
for $j=0,\dots,N-1$, which has a non-trivial solution
$(a_0,\dots,a_{N-1})$. Therefore the function of $\lambda$ defined by
\begin{equation}
    d_N(\lambda) = d_N(\lambda; \lambda_1,\dots,\lambda_{N-1})
    :=
    \left| \begin{matrix}
    \phi^{(N-1)}(\lambda)  & \dots 
                   &\phi'(\lambda)   & \phi(\lambda)   \\
    \phi^{(N-1)}(\lambda_1)& \dots 
                   &\phi'(\lambda_1) & \phi(\lambda_1) \\
    \vdots & \ddots & \vdots & \vdots \\
    \phi^{(N-1)}(\lambda_{N-1}) & \dots 
                   & \phi'(\lambda_{N-1}) & \phi(\lambda_{N-1})
    \end{matrix}\right|
\label{dN(lambda)}
\end{equation}
has a zero at $\lambda= \lambda_0 = c_A - \sum_{i=1}^{N-1}
\lambda_i$. In addition, there are $N-1$ trivial zeros of $d_N(\lambda)$
at $\lambda=\lambda_1,\dots, \lambda_{N-1}$. Hence, from the periodicity
with respect to the lattice $2 \omega_1 \bbZ + 2 \omega_3 \bbZ$ and the
order of poles, it follows that
\begin{equation}
    d_N(\lambda; \lambda_1,\dots,\lambda_{N-1})
    = \tilde d_N(\lambda_1,\dots,\lambda_{N-1})
    \sigma(\lambda)^{-N}
    \sigma\left(\lambda+\sum_{i=1}^{N-1}\lambda_i - u_0\right)
    \prod_{j=1}^{N-1}\sigma(\lambda-\lambda_j),
\label{dN=prod(sigma)}
\end{equation}
where $\tilde d_N$ does not depend on $\lambda$. To determine the
function $\tilde d_N$, we have only to compare the coefficient of
$\lambda^{-N}$ of the Laurent expansion of both sides at $\lambda=0$,
and by induction we have
\begin{equation}
 \begin{split}
    &\left| \begin{matrix}
    \phi^{(N-1)}(\lambda_1)  & \dots 
                   &\phi'(\lambda_1)   & \phi(\lambda_1)   \\
    \phi^{(N-1)}(\lambda_2)& \dots 
                   &\phi'(\lambda_2) & \phi(\lambda_2) \\
    \vdots & \ddots & \vdots & \vdots \\
    \phi^{(N-1)}(\lambda_{N}) & \dots 
                   & \phi'(\lambda_{N}) & \phi(\lambda_{N})
    \end{matrix}\right|
\\
    =&
    \sigma(u_0)^{-1} 
    \left(\prod_{j=1}^{N-1}j! \right) 
    \sigma\left(\sum_{i=1}^{N}\lambda_i - u_0 \right)
    \prod_{i=1}^N \sigma(\lambda_i)^{-N}
    \prod_{1\leqq i < j \leqq N} \sigma(\lambda_i - \lambda_j),
 \end{split}
\end{equation}
from which follows \eqref{det} with $f_n$ defined by \eqref{fn}. In
general, the determinant differs from this case only by a non-zero
constant factor.
\qed

A similar formula with the derivatives of Weierstrass' $\wp$ function
instead of $\phi$ was found in the 19th century. See \cite{bib:Wh-Wa},
p.458.


\end{document}